\documentclass[prl,twocolumn]{revtex4}
\usepackage{epsfig}
\usepackage{amssymb}
\begin{document}

\noindent{\large\bf Comment on ``Triggering Rogue Waves in} 

\vspace{1mm}

\noindent{\large\bf Opposing Currents''}

\vspace{4mm}

\noindent{Victor P. Ruban}

{\small Landau Institute for Theoretical Physics RAS,}

{\small 2 Kosygin Street, 119334 Moscow, Russia}


\vspace{3mm}

\noindent{7 November 2011}

\noindent{===============================}


\vspace{4mm}

The authors of a recent Letter [1] based their study of rogue waves 
in nonuniform currents on a modified nonlinear Schr\"odinger equation
(NLSE; see Eq.(1) in [1]). However, I will show below that equation 
is not correct. It gives wrong solutions even in the first order 
on the supposedly small parameter $U/c_{\rm g}$, where $U(x)$ 
is a current, and $c_{\rm g}=g/(2 \omega)$ [here $\omega$ is 
a mean frequency of a quasi-monochromatic  wave train, and $g$ is
the gravity acceleration]. I also suggest an accurate variant of NLSE 
[see Eq.(\ref{NLSE_Theta}) below], valid in the presence of a large-scale 
nonuniform current under condition $(1+4\omega U/g) \gtrsim 0.2$. 
 
The key point in the derivation is that with non-small 
$U/c_{\rm g}$ we may not assume a globally narrow 
wave-number spectrum, since a local wave-number $k$ essentially 
[up to several times!] varies along $x$ together with $U$ 
[see Eq.(\ref{k}) below]. 
First, we should find monochromatic solutions of the following 
linearized water-wave problem:
$\eta_t=-(U\eta)_x+ {\hat k} \psi$, and $-\psi_t=U\psi_x +g\eta$,
where $\eta(x,t)$ is a vertical displacement of the free water surface 
from a steady-state profile, $\psi(x,t)$ is a surface value of the wave 
velocity potential, and a linear pseudo-differential operator ${\hat k}$ 
multiplies the Fourier image $\psi_k(t)$ by $|k|$. 
Assuming $\eta=\mbox{Re}[Q(x,\omega)\exp(-i\omega t)]$ and
$\psi=\mbox{Re}[P(x,\omega)\exp(-i\omega t)]$,  we have
\begin{equation}
\label{P_omega}
\omega^2 P +i\omega(\partial_x U+U\partial_x)P
-\partial_x U^2 \partial_x P=g{\hat k}P.
\end{equation}
Since $U(x)$ changes over many typical wave-lengths, we can write 
an approximate solution of Eq.(\ref{P_omega}) in the form
$P(x,\omega)\approx\Psi(x,\omega)\exp\Big[i\!\int^x\!\! k(\omega,U)dx\Big]$,
where $\Psi(x,\omega)$ is a slowly varying function, and a positive function 
$k(\omega,U)$ satisfies the local dispersion relation for deep-water waves,
$\omega=Uk+\sqrt{gk}$. In an explicit form, 
\begin{equation}
\label{k}
k(\omega,U)=[g+2\omega U-\sqrt{g^2+4g\omega U}]/(2U^2).
\end{equation}
It is important that with positive $k$ we have 
${\hat k}P\approx-i\partial_x P$. 
Substitution into Eq.(\ref{P_omega}) gives us the equation
$$
\omega (U_x \Psi + 2 U \Psi_x)-2kU(U_x\Psi+U\Psi_x)
-U^2k_x\Psi+g\Psi_x\approx 0.
$$
Multiplying it by $\Psi$ and integrating on $x$, we obtain
$[\omega U-kU^2 +g/2]\Psi^2\approx\mbox{const}$ 
(it is equivalent to the wave action conservation [2]), or 
$\Psi\approx -iC [1+4\omega U/g]^{-1/4}$,
with a complex constant $C$. Since $gQ=i\omega P-UP_x$, we have
$Q(x,\omega)=CM(x,\omega)\exp\left[i\int^x k(\omega,U)dx\right]$, where
$M\approx (k/g)^{1/2}[1+4\omega U/g]^{-1/4}$. What is important, 
at small $U$ the wave amplitude behaves as $M/M_0\approx (1-2\omega U/g)$, 
while in [1] the corresponding factor is  
$\exp[- U/(2c_{\rm g})]\approx (1-\omega U/g)$, which is not correct.

Let us now consider a linear superposition of the mono\-chromatic 
solutions in a narrow frequency range near $\omega$, 
\begin{eqnarray}
\label{envelope}
\eta&=&\mbox{Re}\!\int\! d\xi {\tilde C}(\xi)M(x,\omega+\xi)\,
e^{\{-i(\omega+\xi)t+i\int^x k(\omega+\xi,U)dx\}}\nonumber\\
&\approx&\mbox{Re}\big[ \Theta(x,t)M(x,\omega)\,
e^{\{-i\omega t+i\int^x k(\omega,U)dx\}}\big],
\end{eqnarray}
where $\Theta(x,t)$ is defined by the following integral,
$$
\Theta(x,t)=\int d\xi {\tilde C}(\xi)\,
e^{\{-i\xi t+i\int^x[k(\omega+\xi,U)-k(\omega,U)]dx\}}.
$$
Since the frequency spectrum ${\tilde C}(\xi)$ is concentrated at 
small $\xi$ in a range $\Delta\Omega\ll\omega$, we can expand 
$[k(\omega+\xi,U)-k(\omega,U)]$ in powers of  $\xi$, 
and thus derive a partial-differ\-ential
equation for $\Theta(x,t)$ in the linear regime,
$-i\Theta_x=i k_\omega \Theta_t -(1/2)k_{\omega\omega}\Theta_{tt}+\cdots$.
Here $k_\omega$ and $k_{\omega\omega}$ are partial derivatives of 
$k(\omega,U)$:
$k_\omega=(4\omega/g)[1+\upsilon+(1+\upsilon)^{1/2}]^{-1}$ and
$k_{\omega\omega}=(2/g)(1+\upsilon)^{-3/2}$, 
with $\upsilon\equiv 4\omega U(x)/g$ 
[wave blocking takes place at $\upsilon_*=-1$].

Eq.(\ref{envelope}) means that the wave envelope is 
$\tilde A(x,t)\approx \Theta(x,t)M(x,\omega)$ 
[$\tilde A$ here is not the same as $A$ in [1], but $|\tilde A|=|A|$].
It results in the following estimate for the quantity
$I=(k|A|)(k/\Delta K)$, where $\Delta K\approx k_\omega \Delta\Omega$ 
is a width of the local spectral $k$-distribution,
\begin{equation}
\label{I}
I_\upsilon/I_0=2^{\frac{3}{2}}[1+\upsilon/2+\sqrt{1+\upsilon}]^{-\frac{5}{2}}
(\sqrt{1+\upsilon}+1)(1+\upsilon)^{\frac{1}{4}}.
\end{equation}
Thus, Eq.(8) in [1] should be corrected as written below,
\begin{equation}
A_{\rm max}(\upsilon)M_0/(M_\upsilon A_0)=
1+2\sqrt{1-[\sqrt{2}\varepsilon N I_\upsilon/I_0]^{-2}}.
\end{equation}
However, applicability of formula (\ref{I}) implies 
$(\Delta\Omega)^2\ll 6 k_\omega/k_{\omega\omega\omega}$;
otherwise $\Delta K\not\approx k_\omega \Delta\Omega$. 
Since practically important are values $\Delta\Omega\approx(0.1...0.2)\omega$,
Eq.(\ref{I}) can be good at $\upsilon \gtrsim -0.8$ only.
The same condition arises when we require that the neglected linear
higher-order dispersive terms are small comparatively to
$k_{\omega\omega}\Theta_{tt}$.

To complete our derivation of NLSE with variable coefficients for a weakly 
nonlinear deep-water wave train in a large-scale nonuniform current,
we have in a standard manner to take into account the nonlinear frequency 
shift, which for fixed $k$ is well known to be 
$\delta\omega\approx\sqrt{gk}k^2|A|^2/2$. 
For fixed $\omega$, that corresponds to the nonlinear wave-number shift 
$\delta k\approx -k_\omega \sqrt{gk}k^2|A|^2/2$.
Using the relation $|A|\approx|\Theta| M$, we finally derive
\begin{equation}
\label{NLSE_Theta}
i\Theta_x+i k_\omega \Theta_t -\frac{1}{2}k_{\omega\omega}\Theta_{tt}
-\frac{k_\omega k^3 \sqrt{k}}{2\sqrt{g+4\omega U}}|\Theta|^2 \Theta\approx 0.
\end{equation}

\vspace{5mm}

\centerline{=============================}

[1] M. Onorato, D. Proment, and A. Toffoli, Phys. 

\hspace{3mm} Rev. Lett. {\bf 107}, 184502 (2011).

[2] F. P. Bretherton and C. J. R. Garrett,  Proc. R.

\hspace{3mm} Soc. Lond. A {\bf 302}, 529 (1968).

\end{document}